# A Comparative Study of National Cyber Security Strategies of ten nations

*Adejoke T. Odebade[1], Elhadj Benkhelifa[2]*

[1]*School of Digital Technologies and Art, Staffordshire University, Stoke-on-Trent, UK*

[2]*School of Digital Technologies and Art, Staffordshire University, Stoke-on-Trent, UK*

Corresponding author: *Adejoke T. Odebade (adejoke.odebade@research.staffs.ac.uk)*

## ABSTRACT

This study compares the National Cybersecurity Strategies (NCSSs) of publicly available documents of ten nations across Europe (United Kingdom, France, Lithuania, Estonia, Spain, and Norway), Asia-Pacific (Singapore and Australia), and the American region (the United States of America and Canada). The study observed that there is not a unified understanding of the term "Cybersecurity"; however, a common trajectory of the NCSSs shows that the fight against cybercrime is a joint effort among various stakeholders, hence the need for strong international cooperation. Using a comparative structure and an NCSS framework, the research finds similarities in protecting critical assets, commitment to research and development, and improved national and international collaboration. The study finds that the lack of a unified underlying cybersecurity framework leads to a disparity in the structure and contents of the strategies. The strengths and weaknesses of the NCSSs from the research can benefit countries planning to develop or update their cybersecurity strategies. The study gives recommendations that strategy developers can consider when developing an NCSS.

## 1 Introduction

Cyberspace provides a tremendous opportunity for growth and development (ENISA, 2012). Its effective usage, especially in the Internet of Things, big data, and cloud computing, greatly influences national competitiveness (Min et al., 2015). However, it comes with challenges, such as cyber threats and attacks. Various countries over the past decade have taken steps to address the challenges of cyber threats by developing cybersecurity strategies, enacting cybersecurity laws, and ensuring safeguarding measures to protect customer data (Dedeke & Masterson, 2019). A national-level strategy is crucial to securing cyberspace to ensure prosperity in the digital world (Teoh & Mahmood, 2017). It gives an extensive plan of how an organisation intends to achieve its aim and objectives and make the best use of its unique qualities (Lepori et al., 2013). "Priorities for national cybersecurity strategies will vary country by country. In some countries, the focus may be on protecting critical infrastructure risk, while other countries may focus on protecting intellectual property, and still, others may focus on improving the cybersecurity awareness of newly connected citizens" (Goodwin & Nicholas, p. 2013). These statements show that countries develop their national cybersecurity strategy based on their understanding and perception of cybersecurity. What constitutes a significant risk for one country may not apply to another country. Therefore, governments will put together strategies to best protect their economy and citizens.

Structuring a National Cyber Security Strategy (NCSS) can be in areas such as investment in research and development, awareness and training, collaboration and information sharing, and partnership within government organisations, depending on the need and perception of the nation (Min et al., 2015). It is fundamental to protecting cyberspace from malicious attackers and providing a safe environment for the digital economy to thrive (Teoh & Mahmood, 2017; Shafqat & Masood, 2016); therefore, effective sustenance of cybersecurity is achievable through an enforceable national strategy (Ghernouti-Hélie, 2010). Consequently, a national strategy's goals, objectives, scope, and priorities must be defined to foster partnerships among stakeholders and communicate a nation's objectives to other countries and stakeholders (Luiijf et al., 2013; Sabillon et al., 2016). In developing an effective cybersecurity strategy, considerations of both national and international needs are paramount because cybercrime is a national as well as a global issue, necessitating the need for international collaboration and the development of national and international strategies to combat cybercrime (Goodwin & Nicholas, p. 2013; Ghernouti-Hélie, 2010). NCSS should be current, adequate, and suitable to avoid putting ICT (Information and Communications Technology) and the lives of citizens at risk (Mori & Goto, 2018).

The paper is organised as follows. Following the introduction, we present related work in the domain, followed by the method and approach. Next, we present the results from the review and analysis, after which we present the discussion. The next section contains the framework mapping of the NCSS, and the last two sections explain the conclusion and recommendations.

## 2 Related Work

NCSS, as described by (ENISA, 2016,) is a tool the government use in setting guidelines, principles, measures, and objectives to mitigate cyber security risks. The principles must be clear to aid decision-making in identifying, managing, or mitigating risks associated with cyber security (Goodwin & Nicholas, 2013). Previous studies have analysed various NCSSs mainly through a comparative approach. A cross-comparison survey based on five focal points compared fourteen NCSS [10], and common themes identified were legal issues, public-private partnerships, and early-warning methods. Differences identified range from technical to complex national security issues. Luiijf et al. (2011) analysed and compared ten NCSS based on nine topics and made thirteen observations NCSS should consider. Luiijfet et al. (2013) further explored and compared nineteen NCSSs of eighteen nations which looked at the similarities and differences in the definition of cybersecurity, general information, visions, objectives and principles, tactical/operational level action plans, and International NCSS development. The research found that the strategies were unclear about their relationship with other national and international policies. Further differences in military defence, national security, or economic approaches were also observed. Lehto (2013) carried out research that analysed the threats, definitions, and objectives of eight nations using the fivefold classification model of cybercrime, cyber espionage, cyberactivism, cyber terrorism and cyber warfare. The research found significant variance in the depth and scope of the cybersecurity strategies and the differences in the emphasis and priorities placed on the private sector, citizen perspective and public administration. Tatar et al. (2014) examined and compared the NCSS of six nations and noted variations in the nations' focus ranging from the economic impact of cyber incidents, cyber-attack prevention on critical sectors, military involvement and sustaining the society. A lack of methodology in evaluating the strategies was noted.

An international comparative study on cyber security strategy by Min et al. (2015) was carried out with an emphasis on public-private partnerships and how the institutional framework of the partnership was established. The study noted that public-private partnerships are strengthened under the Government's authority to respond effectively to cyber security accidents. A paradigm shift in government intervention in the private sector was observed, stating a change from voluntary to enforced self-regulation. The scope of the research was, however, limited to three nations. The NCSS review of fifty-four countries by (Azmi et al., 2016) focused on national security, jurisprudence, and politics. It was observed that the NCSS supports national security considering the benefits and risks associated with internet usage. A comparative analysis of twenty countries based on technical, policy-related, legal, and operational measures was carried out by Shafqat and Masood (2016), using the ITU's cyber security ranking. The research found that although there were common aims and objectives, there were also considerable differences in the scope and approach of the strategies. The study gave detailed recommendations to consider when developing an NCSS and concluded that three of the nations' strategies were better than the rest. Kolini and Janczewski (2017) analysed sixty NCSS, and the authors, using a hierarchical clustering method, noticed similarities between NCSS developed by EU or NATO members. A common focus of the majority of the NCSS was critical infrastructure protection, public-private partnership and defending the IT system of the wider society.

## 3 Method and Approach

This study examined the literature of publicly available documents. The framework for the comparison was based on Global Cybersecurity Index (ITU, 2019) and Luiijf et al. (2013). In addition, the study used the ITU guide (2018) to provide extensive research into the various aspects of NCSS. The data set includes twenty NCSSs published in English across ten nations. The analysis uses tables as a quick reference to the key elements of the strategies. The main issues explored are:

- Chronological development
- Definition of Cyber security
- Strategic vision and timeframe
- Strategic objectives
- Guiding principles
- Responsible agency
- Method / Methodology
- Threat subjects and malicious threat actor objectives
- Response team
- Child online protection mechanisms
- Capacity-building measures: public awareness campaigns, training and educational programs, research and developmental programs and incentive mechanisms.
- Cooperation measures: international cooperation and public-private partnership.

Furthermore, for this research, the NCSSs for this study were selected from the 2018 normalised score of the GCI of the most committed countries to cybersecurity (ITU, 2019). Six countries are within the Europe region (United Kingdom, France, Lithuania, Estonia, Spain and Norway), two from the Asia-Pacific region (Singapore and Australia) and two from the Americas region (the United States of America and Canada). Although Malaysia is 8th in the ranking, it was excluded from the comparative analysis due to insufficient information. Table 1 shows the GCI ranking of the selected countries and abbreviations used:

Table 1: Selected countries based on Global Cybersecurity Index (GCI)

| Ranking | Country | Abbreviation |
|---|---|---|
| 1 | United Kingdom | UK |
| 2 | United States of America | USA |
| 3 | France | FRA |
| 4 | Lithuania | LTU |
| 5 | Estonia | EST |
| 6 | Singapore | SG |
| 7 | Spain | ESP |
| 9 | Norway | NOR |
| 10 | Canada | CAN |
| 11 | Australia | AUS |

## 4. Results from review and analysis
The key elements explored are discussed below.

### 4.1 Chronological development of the cyber security strategy

Sponsored cyber-attacks by nation-states after 2008 led to an increase in the development of cybersecurity strategies (Shafqat & Masood, 2016). Of the ten NCSS, the most recent are Lithuania (Ministry of National Defence, 2018), Estonia (Ministry of Economic Affairs and Communications. 2019), USA (The White House, 2018), Norway (Norwegian Ministries, 2019), Canada (Public Safety, 2018) and Spain (Supreme Council, 2019). USA and Norway were the first countries to produce their strategies in 2003 (Norwegian Ministries, 2003; The White House, 2003). It has taken the USA fifteen years to create a second strategy (The White House, 2018) compared to Norway, which has produced four strategies (Norwegian Ministries, 2003; 2007; 2012; 2019) over sixteen years. In 2009, the USA published a Strategy Review (The White House, 2009) and the Defence Strategy in 2015 (Department of Defence, 2015), possibly explaining the gap in the US strategies. Table 2 gives a chronology of the ten NCSSs and some general information.

Estonia and Lithuania joined the European Union (EU) in 2004 (Estonian Ministry of Foreign Affairs, 2009; Ministry of Foreign Affairs of the Republic of Lithuania, 2006) but Lithuania has only one national cyber security strategy (Ministry of National Defence, 2018). As noted in the strategy, their lack of fully established cybersecurity research and education is due to their late entry into the EU. Estonia, however, produced its first strategy four years after joining the EU and has three strategies in total (Ministry of Defence, 2008; Ministry of Economic Affairs and Communications, 2014; 2019).

Table 2: Chronological development and general information of the NCSSs.

| Nations | LTU | EST | USA | FRA | SG | NOR | CAN | AUS | ESP | UK |
|---|---|---|---|---|---|---|---|---|---|---|
| Year | 2018 | 2019 | 2018 | 2015 | 2016 | 2019 | 2018 | 2016 | 2019 | 2016 |
| Period | 5 | 3 | NS | NS | NS | NS | 5 | 4 | NS | 5 |
| Total number of NCSS | 1 | 3 | 2 | 2 | 1 | 4 | 2 | 2 | 2 | 3 |
| Timeline of strategies | 2018 | 2008 2014 2019 | 2003 2018 | 2011 2015 | 2016 | 2003 2007 2012 2019 | 2010 2018 | 2009 2016 | 2013 2019 | 2009 2011 2016 |

The development of NCSS from table 2 gained momentum around 2008; however, there is inconsistency in the time interval of subsequent ones. The average span of the NCSS is about 4-6 years before the development of a new one which poses a risk considering the emerging threats in cyber security. Although developing a new strategy (for instance, every two years) is hard to achieve, a regular review of the existing strategy is essential to ensure that it fits its purpose and meets set objectives.

**4.2 Definition of cyber security**

There is no universally acceptable definition of cybersecurity (Min et al., 2015; Shafqat & Masood, 2016; Luiijf et al., 2013). Hence, the lack of a unified definition may confuse nations when deliberating global cyber issues (Luiijf et al., 2013). The term (if defined) is based on the frame and needs of various stakeholders; therefore, there are diverse definitions of cybersecurity, which is sometimes synonymous with digital security. According to the International Organisation for Standardization (ISO), cybersecurity is defined as "the preservation of confidentiality, integrity and availability of information in Cyberspace (ISO, 2012). International Telecommunications Union (ITU) defines Cyber security as: "the collection of tools, policies, security concepts, security safeguards, guidelines, risk management approaches, actions, training, best practices, assurance and technologies that can be used to protect the cyber environment and organization and user's assets" (ITU, p. 2009). Table 3 shows the varied definitions of cybersecurity across the strategies.

Spain (Supreme Council, 2019), Lithuania (Ministry of National Defence, 2018), USA (The White House, 2018), Australia (Australian Government, 2016), France (the Republic of France, 2015) and Singapore (Cyber Security Agency of Singapore, 2016) used the term cybersecurity with no definition of the term in their strategies. Although Estonia's most recent strategy (Ministry of Economic Affairs and Communications, 2019) states that terminologies were in the appendix, there was no appendix in the document publicly available; it, however, implicitly defines the term. Norway (Norwegian Ministries, 2019), Canada (Public Safety, 2018) and the UK (HM, Government, 2016) explicitly define cyber security in the most current strategies. Norway used and defined the term information security in its 2007 strategy (Norwegian Ministries, 2007); however, the 2012 and 2019 strategies (Norwegian Ministries, 2012; 2019) defined cybersecurity. The 2019 definition includes technology and not only connected items to the internet. Canada's initial definition in the 2010 strategy (Government of Canada, 2010) was inferential from the definition of cyber-attack compared to the 2018 (Public Safety, 2018) with a clear definition; UK's definition in 2009 strategy (Cabinet Office, 2009) was quite specific to the UK. However, the latest strategy (HM, Government, 2016) gives a far-reaching description of the term. Australia and France presented a clear definition of cybersecurity in their first strategy (Australian Government, 2009; ANSSI, 2009) but failed to define it in the current strategy.

Even though the definitions across the countries differ, there has been an improvement in the understanding of the concept. Norway defines cyber security as protecting 'everything' that is vulnerable. The definition is quite vague, as what 'everything' entails is unclear. Additionally, Norway uses the term synonymously with ICT security and digital security; however, there are differences in the definitions. ICT security, as defined by Von Solms & Van Niekerk (2013), is "the protection of the actual technology-based systems on which information is commonly stored and/or transmitted" (p. 98). Although the UK and Canada both described the term by stating what is being protected, the UK further includes non-digital services, recognising that damage caused can be deliberate or accidental.

In an improved definition of the term cyber security, Schatz et al. (2017) carried out various semantic and lexical analyses on definitions from authoritative sources – Industry, Government and nation-states, and Academia. Upon research, they defined cyber security as "The approach and actions associated with security risk management processes followed by organizations and states to protect confidentiality, integrity and availability of data and assets used in cyber space. The concept includes guidelines, policies and collections of safeguards, technologies, tools, and training to provide the best protection for the state of the cyber environment and its users" (p. 66). This descriptive definition, though lengthy, encapsulates the core pillars of cyber security: people, processes, and technology. Comparing the above definition to those provided in the NCSSs, it is apparent that nations have not fully embraced the concept.

Nations have developed their strategies considering social, political, and other prevalent conditions and, as such, will not have a single strategy that will suit them all. However, the lack of a unified definition of "cyber security" that brings them together should be addressed. Organisations across countries collaborate and work in partnership to address cybersecurity challenges, yet there is no joint agreement and understanding of the term. While countries may not be on the same spectrum in terms of cybersecurity, a consensus on the term is a good starting point such that upcoming nations in this domain can have a platform to build on.

Table 3: Definitions of Cybersecurity of the NCSS

| Country | Year | Definition |
|---------|------|------------|
| NOR | 2003 | Indeterministic as the document is in Norwegian |
| | 2007 | References to information security: protection against breaches of confidentiality, integrity or availability of the information that is being processed in a system, or the protection of information systems and networks in themselves. |
| | 2012 | Protection of data and systems connected to the Internet |
| | 2019 | Cyber security has to do with protecting "everything" that is vulnerable because it is connected to or otherwise dependent on information and communication technology. The term is used synonymously with the terms "ICT security" and "digital security". |
| CAN | 2010 | The appropriate level of response and/or mitigation measures to cyber-attacks- the unintentional or unauthorized access, use, manipulation, interruption or destruction (via electronic means) of electronic information and/or the electronic and physical infrastructure used to process, communicate and/or store that information. |
| | 2018 | The protection of digital information, as well as the integrity of the infrastructure housing and transmitting digital information. More specifically, cyber security includes the body of technologies, processes, practices and response and mitigation measures designed to protect networks, computers, programs and data from attack, damage or unauthorized access so as to ensure confidentiality, integrity and availability. |
| AUS | 2009 | Measures relating to the confidentiality, availability and integrity of information that is processed, stored and communicated by electronic or similar means. |
| | 2016 | Not Defined |
| ESP | 2013, 2019 | Not Defined |
| UK | 2009 | Cyber security embraces both the protection of UK interests in cyberspace and also the pursuit of wider UK security policy through exploitation of the many opportunities that cyberspace offers. |
| | 2011 | Not Defined |
| | 2016 | The protection of information systems (hardware, software and associated infrastructure), the data on them, and the services they provide, from unauthorised access, harm or misuse. This includes harm caused intentionally by the operator of the system, or accidentally, as a result of failing to follow security procedures. |
| LTH | 2018 | Not Defined |

| | | | |
|---|---|---|---|
| **EST** | **2008, 2014,** | Protecting digital society and the way of life as a whole | |
| | **2019** | Protecting digital society and the way of life as a whole | |
| **USA** | **2003, 2018** | Not Defined | |
| **FRA** | **2011** | The desired state of an information system in which it can resist events from cyberspace likely to compromise the availability, integrity or confidentiality of the data stored, processed or transmitted and of the related services that these systems offer or make accessible. | |
| | **2015** | Not Defined | |
| **SG** | **2016** | Not Defined | |

### 4.3 Strategic vision and time frame

A national strategy provides a platform for a country to communicate its intention to other countries and interested parties Luiijf (2013) and outlines a vision which describes an organisation's intended future state (Orhan et al., 2014; Greene, 2019). A vision is described as the significance of the work (Tvorik & Mcgivern, 1996), a comprehensive notion upon which other notions are absorbed (Collins & Porras, 2008) and a sense of direction (Jones, 2010). It defines the reason for existence and what an organisation should achieve to succeed (Greene, 2019). Table 4 shows that seven nations explicitly provided a vision for the NCSS; two countries implicitly described the goal of their strategy, while one nation did not define its vision.

Table 4: Vision and Time frame of NCSS

| Country | | Vision | Time frame (years) |
|---|---|---|---|
| CAN | Explicit | Security and prosperity in the digital age | 5 |
| SG | Explicit | To create a resilient and trusted environment | None |
| NOR | Explicit | It is safe to use digital services. Private Individuals and companies have confidence in national security, and trust that the welfare and democratic rights of the individual are being safeguarded in a digitalised society. | None |
| US | Explicit | An open, interoperable, reliable, and secure Internet | None |
| ESP | Explicit | General goal in line with the National Security Strategy is guarantee secure, reliable use of cyberspace, protecting citizens' rights and freedoms and promoting socio-economic progress. | None |
| UK | Explicit | The UK is secure and resilient to cyber threats, prosperous and confident in the digital world | 5 |
| LTH | Implicit | Provide the Lithuanian people with the opportunity to explore the potential of information and communications technology (ICT) by identifying cyber incidents timely and effectively, by preventing cyber incidents and their recurrence, and by managing the impact of cybersecurity breaches | 5 |
| EST | Explicit | The most resilient digital society | 3 |
| AUS | Explicit | Open, free and secure Internet | 4 |
| FRA | | Not Defined | None |

A common theme across the visions of the NCSSs is security and resilience. Canada and UK focused on security and economic prosperity in the digital world, while Singapore and Estonia focused on resilience. Australia and USA focused on open and secure Internet. At the same time, Norway sets itself as one of the leading nations in the digital age with a focus on security and citizens' trust. The emphasis of the Spanish strategy is on security and citizens' economic growth, while Lithuania's is on awareness and resilience. Estonia's distinctive vision of "the most resilient digital society" sets the country a world leader in cybersecurity. This may be associated with the ability to cope with cyber-attacks because of the 2007 vicious cyber-attack on the Estonian economy (Herzog, 2011). Cyber risk awareness and a highly skilled workforce are crucial to achieving the vision of Estonia's strategy. Although the US vision states an open, interoperable, reliable, and secure Internet, a report from Freedom House showed a decline in internet freedom in the USA (Freedom House, 2018). A vision of an NCSS should be time-bound so that the strategy's objectives can be broken down into manageable tasks. Only five countries (CAN, UK, LTH, EST and AUS) provided a timeframe for their NCSS. The vision of a strategy must be clear and measurable; however, most of the vision statements projected in these strategies though concise, are rather vague.

### 4.4 Strategic objectives

A vision can be broken down into strategic objectives (Singh, 2017) and should be SMART: Specific, Measurable, Assignable, Realistic, and Time-related (Doran, 1981). Common goals across the NCSS are building a cyber security culture through education, fostering international cooperation, promoting research and development, promoting cyber awareness, and creating an environment of trust in cyberspace for citizens, businesses, and Government to operate.

Table 5: Goals and Objectives of the NCSSs

| Country | Goals / Objectives |
|---------|-------------------|
| CAN | 1. Security and Resilience<br>2. Cyber Innovation<br>3. Leadership and Collaboration |
| SG | 1. Building a resilient infrastructure<br>2. Creating a safer cyberspace<br>3. Developing a vibrant cybersecurity ecosystem<br>4. Strengthening international relationships |
| NOR | 1. Norwegian companies digitalise in a secure and trustworthy manner, and are able to protect themselves against cyber incidents<br>2. Critical societal functions are supported by a robust and reliable digital infrastructure<br>3. Improved cyber security competence is aligned with the needs of society<br>4. Society has improved ability to detect and handle cyber attacks<br>5. The police have strengthened their ability to prevent and combat cyber crime |
| USA | 1. Protect the American People, the Homeland, and the American Way of Life<br>2. Promote American Prosperity<br>3. Preserve Peace through Strength<br>4. Advance American Influence. |
| ESP | 1. Security and resilience of information and communication networks and systems for the public sector and essential services<br>2. Secure and reliable use of cyberspace to ward off illicit or malicious use<br>3. Protecting the business and social ecosystem and citizens<br>4. Culture and commitment to cybersecurity and strengthening human and technological skills<br>5. International cyberspace security. |
| UK | 1. Defend our cyberspace<br>2. Deter our adversaries<br>3. Develop our capabilities. |
| LTH | 1. To strengthen the cybersecurity of the country and the development of cyber defence capabilities<br>2. To ensure prevention and investigation of criminal offences in cyberspace<br>3. To promote cybersecurity culture and development of innovation<br>4. To strengthen close cooperation between private and public sectors<br>5. To enhance international cooperation and ensure the fulfilment of international obligations in the field of cybersecurity. |

| EST | 1. A sustainable digital society |
| | 2. Cybersecurity industry, research and development |
| | 3. A leading international contributor |
| | 4. A cyber-literate society. |
| AUS | 1. A national cyber partnership |
| | 2. Strong cyber defences |
| | 3. Global responsibility and influence |
| | 4. Growth and innovation |
| | 5. A cyber smart nation. |
| FRA | 1. Fundamental interests defence and security of State information systems and critical infrastructures, major cybersecurity crisis |
| | 2. Digital trust, privacy, personal data, cyber malevolence |
| | 3. Awareness raising, initial training, continuing education |
| | 4. Environment of digital technology businesses, industrial policy, export and internationalisation |
| | 5. Europe, digital strategic autonomy, cyberspace stability |

The UK strategy aims to report annually on the progress made in achieving its objectives. However, there is no documentation to show that annual reports have been made three years into the strategy. A UK report by the National Audit Office (National Audit Office, 2019) reveals that 1 out of the 12 strategic outcomes has high confidence that it will be achieved by 2021, 4 is accessed as moderate confidence, 6 as low confidence and the last strategic outcome was not analysed. This suggests that the prospects of the UK achieving its vision for 2021 are uncertain. Australia also mentioned that annual updates on progress made would be published, however, after the first annual update in 2017 (Australian Government, 2017), no further updates were published. Canada committed to reporting on the progress made, but it failed to give a timeline; hence, it is uncertain if progress reports will be published soon.

Table 5 above outlines the strategic objectives of the ten nations. Although common themes span various objectives, each strategy has its unique goals. For instance, Norway's strategy intends that individuals understand the risk associated with using technology and can use it safely to tackle future challenges related to the overarching changes in the digitalisation of Norwegian society. In France, the bedrock for the national strategy is training and international cooperation to promote the economy, citizens, and national values. Estonia's strategy is to ensure that critical infrastructures are resilient to cyberattacks. Lithuania's strategy focuses on resilience and raising public awareness, while the UK focuses on confidence and resilience. Australia's strategy aims to promote cyber security hygiene and raise awareness. Although the UK has classified its objectives into three themes, there are thirteen strategic outcomes in the strategy.

### 4.5 Guiding principles
Values justify our priorities and ethical judgements, and different people have different ethics (Kelly et al., 2015). Table 6 shows the guiding principles that underpin the NCSS of four nations. Norway refers to active cooperation as fundamental to tackling cyber security challenges, while Estonia relates to security, openness, and freedom as essential to cyberspace. The UK's principles are very comprehensive, some of which are shown in the table focusing on protection, accountability, and cooperation, while the Spanish principles are efficiency, resilience, and anticipation. France, Lithuania, Australia, the USA, Singapore, and Canada's strategies did not include guiding principles.

Table 6: Guiding principles

| NOR | 1. The authorities and the business community work together to identify and discuss cyber security challenges and to exchange experiences about them. |
| | 2. This cooperation should carry obligations for both parties and be based on transparency, trust and mutuality. |
| | 3. The authorities contribute to establishing a business community where cyber security services are in demand, developed and provided. |
| | 4. When building up national capacity in cyber security, it should be facilitated for inclusion of capabilities from the business community. |
| EST | 1. We consider the protection and promotion of fundamental rights and freedoms as important in cyberspace as in the physical environment. |

| | | |
|---|---|---|
| | 2. | We see cybersecurity as an enabler and amplifier of Estonia's rapid digital development, which is the basis for Estonia's socio-economic growth. Security must support innovation, and innovation must support security. |
| | 3. | We recognise the security assurance of cryptographic solutions to be of unique importance for Estonia as it is the foundation of our digital ecosystem. |
| | 4. | We consider transparency and public trust to be fundamental for digital society. Therefore, we commit to adhere to the principle of open communication |
| UK | 1. | our actions and policies will be driven by the need to both protect our people and enhance our prosperity. |
| | 2. | we will treat a cyber-attack on the UK as seriously as we would an equivalent conventional attack and we will defend ourselves as necessary. |
| | 3. | we will rigorously protect and promote our core values. These include democracy, the rule of law, liberty; open and accountable governments and institutions; human rights; and freedom of expression. |
| | 4. | the Government will meet its responsibilities and lead the national response, but businesses, organizations and individual citizens have a responsibility to take reasonable steps to protect themselves online and ensure they are resilient and able to continue operating in the event of an incident; |
| | 5. | responsibility for the security of organisations across the public sector, including cyber security and the protection of online data and services, lies with respective Ministers, Permanent Secretaries and Management Boards. |
| | 6. | we will work closely with those countries that share our views and with whom our security overlaps, recognising that cyber threats know no borders. We will also work broadly across the range of international partners to influence the wider community, acknowledging the value of broad coalitions; and |
| ESP | 1. | Action Unit: Any response to an incident in the cybersecurity field that might implicate different State agents will be strengthened if it is coherent, coordinated and resolved quickly and effectively. These qualities can be attained through careful preparation and appropriate organization of the State's action unit. |
| | 2. | Anticipation: The specific nature of cyberspace and the players involved requires anticipation mechanisms in specialised organisations to guide State action in crisis situations. |
| | 3. | Efficiency: Cybersecurity requires the use of high-value, multi-offer systems featuring a high-level technology that are associated with very demanding needs and high costs derived from their development, purchase and operation. |
| | 4. | Resilience: is a fundamental feature for systems and critical infrastructures. The State must ensure availability of elements considered to be essential for the nation, improving their protection against cyber threats. |

**4.6 Responsible agency implementing strategy**

A responsible agency that takes the lead in coordinating the strategy should be identified and empowered to ensure its successful implementation. A responsible agency can include advisory councils, permanent committees, multidisciplinary groups or Government officials and the roles and responsibilities of stakeholders and their inter-relationships should be well defined (ITU, 2019). Canada did not give a clear structure as to the roles and responsibilities of stakeholders within the Government and their inter-relationship, however, efforts will be made by the Government to foster collaboration and provide clear guidance. In Singapore, the Cyber Security Agency (CSA) is the central point for cyber security issues and has been empowered to carry out activities such as enforcing regulations and policies. Governmental bodies coordinate cyber security in Norway and the USA, each having its responsibilities; hence strong collaboration is key to a successful implementation of the strategy. The National Security Council in Spain takes the lead. It is supported by other governmental bodies, while the minister for the cabinet office in the UK takes the lead. Still, a strong partnership with devolved administrations is paramount to the strategy's success. In Lithuania, the Ministry of National Defence and the National Cyber Security Centre (NCSC) and other governing bodies take the lead in implementing the strategy. The Ministry of Economic Affairs and Communications in Estonian takes the lead in coordinating the planning of cybersecurity policy and coordinating the strategy implementation along with other government ministries and the cybersecurity council. In contrast, Australia's special adviser on cyber security takes the lead responsibility for implementing the strategy. In France, the agency responsible for implementing the strategy is not clear, however, it is inferred that the Government and elected officials will take the lead.

NCSS should identify who takes the lead, and their responsibilities should be clear to avoid any conflict of interest with other organisations.

### 4.7 Method and Methodology in preparing a strategy

The critical infrastructures across the countries lie in both the private and public sectors, and as such, both parties should be involved in producing the strategy for maximum cooperation. The strategies were unanimous that Government alone cannot bear cyber security responsibilities and should be a joint effort of all stakeholders. While the UK, Singapore, USA, Lithuania, France and Australia failed to state the method in preparing the strategy and the level of stakeholder involvement, Canada, however, includes public consultation with citizens and key stakeholders, engagement within the federal government cyber security community and an evaluation of the 2010 strategy. Likewise, Norway adopts an all-inclusive approach involving stakeholders from both the private and public sectors through workshops and conferences. Spain involved government ministries, academia, businesses, experts and professionals, while Estonia involved the public sector, academia, providers of critical infrastructures, and businesses, considering lessons learnt from previous strategies. The strategies, however, did not state the methodology.

### 4.8 Threat subjects and malicious threat actor objectives

All the NCSSs recognise the devastating impact cyber threats can have, especially on essential services of the country. Some of the strategies were explicit about their threat actors, while some generalised their actors. Norway pinpoints that foreign activities and cybercrime formed the main cyber threats to their society in 2018, while France avers that over the years, many states have been a threat to their nation, drawing attention to the terrorist attack suffered in January 2015 which led to the setup of an Islamic radicalisation information platform as a means to inform the public on the risks related to such attacks. Lithuania notes that foreign countries pose a threat to their national security, with the National threat assessment (State Security Department of the Republic of Lithuania, 2019) affirming that countries such as Russia and China are a threat to their national security. Estonia states that cyber threat does not respect national boundaries, while Singapore states cyber espionage as one of the major challenges. The UK extensively discussed the various categories of threats it faces, some within the UK. Others are from South Asia, West Africa, and mainly Russian-language organised criminal groups (OCG) in Eastern Europe, where most extreme cybercrimes are carried out. The USA was explicit in some of the threat actors, such as Russia, Iran, North Korea and Chile, while Spain did not refer to specific threat actors but devoted a chapter to cyber threats and actions in cyberspace. Australia notes that severe and organised criminal syndicates and foreign adversaries threaten their interests. At the same time, Canada states that terrorist organisations, state-sponsored actors, and nation-states are a threat to their interests.

Identifying and understanding the threats vectors is vital; however, introducing the concept of risk will minimise threat and vulnerability, where Risk is commonly defined as Threat x Vulnerability x Consequence Chabinsky (2010). Cybersecurity strategy developers should consider completing a Cybersecurity Vectors and Risk chart by applying risk analysis to threat vectors: supply chain and vendor access, remote access, proximity access, and insider access (Chabinsky, 2010). Fig 1 below has been conceptualised from the Chabinsky (2010) risk chart

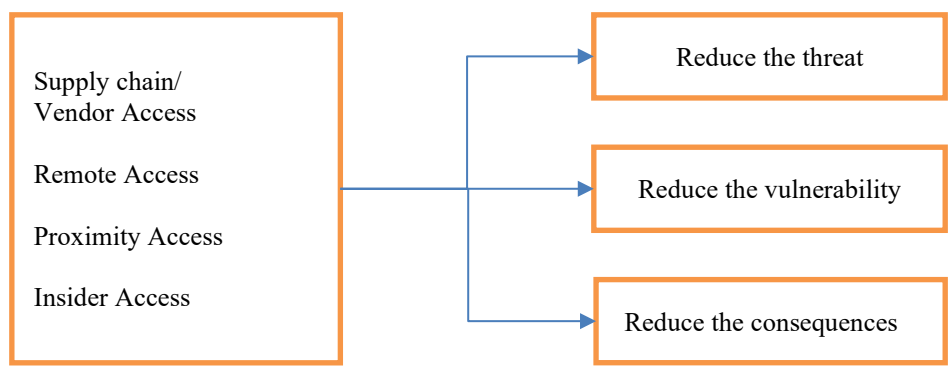

Figure 1 Cybersecurity Vectors and Risk Chart, Chabinsky (2010).

**4.9 Response team**

A CERT (Computer Emergency Response Team) or CIRT (Computer Incident Response Team) is a group of information security experts responsible for managing cyber incidents (Grobler, 2010). In a cyber security incident, it is vital to identify and respond to the incident because the speed in recognising, analysing and responding to the incident determines the level of damage (Ruefle et al., 2014). All the countries except the USA referred to a response team. As mentioned by France, one area of concern is that there is no clear guidance or assistance provided to the public, SMEs, and other stakeholders who are victims of cyber-attacks; however, a national system will be set up to provide support to victims of cyberattacks. It is not sure if this has been achieved. Table 7 shows some of the varied responsibilities of the response teams.

Table 7: Response team

| Country | Response Team | Responsibility |
|---------|---------------|----------------|
| CAN | Cyber Incident Response Centre (CCIRC) | • Provides cyber threat warnings to organisations. |
| SG | Cyber Security Agency | • Central agency for cyber security<br>• Develop and enforce cyber security regulations<br>• Foster collaboration among stakeholders<br>• Protect essential services |
| NOR | National Security Authority (NSM) | • Responsible for coordinating computer attacks on critical infrastructures.<br>• Runs the national response function for severe computer attacks on critical infrastructure (NorCERT) and the national warning system for digital infrastructure (VDI). |
| US | Not Specified | |
| ESP | National CSIRT | • Work with regional and private CSIRTs, to encourage initiatives that help meet national strategic goals. |
| U | NCSC | • Provides advice on threat intelligence<br>• Foster collaboration among various stakeholders<br>• Incident management capabilities |
| LTH | CERT-LT | • Organises cyber incident management. |
| EST | CERT 24/7 | • Monitoring and incident resolution capability. |
| AUS | CERT Australia | • Advises on cyber security threats to Australia's critical infrastructure owners and operators.<br>• Works directly with other computer emergency response teams around the world.<br>• Convenes National and Regional Information Exchanges with businesses. |
| FRA | National Cybersecurity Agency (ANSSI) | • Deals with cybersecurity incidents that affect administrations and operations of vital importance. |

**4.10 Child online protection mechanisms**

The Internet can be a risky and harmful place for children even though it provides an environment where they can also learn (Livingstone et al., 2014). This section looks at any initiative or program to help children stay safe online and provide information and support to parents and wards on how to protect children online. The UK, Canada, Norway, France, Singapore, Spain and the USA did not refer to any child online protection services or initiatives. Estonia mentions the Violence Prevention Strategy for 2015-2020 for children and teens to ensure they safely use the media and are aware of dangers online. There is also the Children and Families Development Plan 2012-2020, a service that provides advice and guidance to parents and a means to report illegal online activities. Australia set up the Office of the Children's eSafety Commissioner to equip children and young people with information to help them stay safe online and provide help with cyberbullying. Lithuania did not mention any child protection initiatives in operation but noted that the EU's directive on sexual abuse and exploitation of children had been transposed into national legislation. Although countries have initiatives and programs tailored toward their citizens, the strategy should include sensitive issues such as the risk faced by children and an overview of how the strategy will address such concerns.

**4.11 Capacity building**

Despite various political and socioeconomic implications, a common approach in addressing cybersecurity has been through technology (ITU, 2019). With the increase in cybercrime in the 21st century, it is essential to understand and find ways to hinder the rise in cyber-attacks (Gritzalis & Tejay, 2013). Both security professionals and regular users must be informed about current cybercrimes and attacks and receive guidance on cyber security which can be achieved through awareness and training (Aloul, 2011; Reid & Niekerk, 2014). Awareness is the bedrock of security culture (Williams, 2009), and the main aim of awareness in cybersecurity is to effect changes in online behaviour, with security being the focal point (Korpela, 2015). Accordingly, awareness programs enable users to identify IT security concerns while training aims to produce suitable and required IT skills and abilities. The fight against cybercrime is a collaborative effort and cannot be achieved by an individual or organisation. Hence, various stakeholders (Governments, private sectors, and national and international bodies) must work in partnership to fight cybercrime as security problems can occur inside and outside an organisation (Reid & Niekerk, 2014). There is a consensus that capacity building is key to tackling the challenges of cyber security, and four areas were looked at as discussed below:

*Public awareness campaigns*: public awareness campaigns look at the awareness to promote cyber safety. A basic level of knowledge and cyber security awareness is required when using the Internet and ICT devices (Reid & Niekerk, 2014; Kritzinger & Von Solms, 2010). All the NCSSs agreed to the need to intensify efforts to promote safe cyber behaviour online as there is still insufficient awareness of the risks of cyber threats. No specific actions were mentioned in Norway, Lithuania, USA and Spain but the UK strategy highlights specific actions such as the 10 steps guidance which has partly helped in improving cyber hygiene over the past few years. Other initiatives like the cyber aware campaign and cyber essentials scheme have also been designed to help individuals and organisations protect themselves from threats and develop cyber hygiene. Although there has been an increase in the awareness of these initiatives in the UK, most businesses however are still not aware of Government initiatives (Department for Digital, Culture, M&S, 2019). Table 8 shows some initiatives across the strategies.

Table 8: Awareness campaign measures

| Country | Key Actions |
|---------|-------------|
| CAN | • Get Cyber Safe Campaign from efforts made under the 2010 strategy<br>• Smart Cities Challenge initiative to improve the quality of life through digital technologies. |
| SG | • Publicity campaigns through media to send prevention messages to the public.<br>• The Public Cyber-Outreach & Resilience Programme (PCORP) initiative to encourage good cyber hygiene practices.<br>• Scam Alert website, Cybersecurity Awareness Campaign, GoSafeOnline, Total Defence and Let's Stand Together campaigns to provide public information on online protection and promote the importance of cybersecurity. |
| NOR, LTH, US, ESP | • Campaigns, initiatives and awareness programs not mentioned although efforts are being made to improve awareness campaigns. |
| UK | • 10 steps to cyber security<br>• Cyber Aware campaign and Cyber Essentials initiatives to improve cyber hygiene standards. |
| EST | • Overlap of activities and information fragmentation among projects without a central channel for the public to access information.<br>• RIA now takes on the leading role in promoting cyber hygiene and raising public awareness. |
| AUS | • Joint public-private awareness initiatives and education campaigns will be improved to improve awareness of cyber security.<br>• Australian Cybercrime Online Reporting Network (ACORN)<br>• Stay Smart Online and SCAMwatch. |
| FRA | • Existing operations such as Internet Licence will be improved to enable sixth-year students to receive advice on Internet browsing.<br>• The "A digital education for all" portal will be enhanced<br>• Cybersecurity training for Information and system engineers in the public sector as part of recruitment program. |

*Training and educational programmes:* the prevalent use of computers and its role in promoting economic prosperity has become a necessity in the global Internet economy (Boulton & Csizmadia, 2018). However, users can be ignorant of the danger posed by using the computer to access Internet and are easy targets for cyber criminals and must therefore be trained and educated to tackle the challenges of cyber threats (Aloul, 2011). This section looks at future or current actions in place to promote cyber security through education and training. All the strategies acknowledge the lack of expertise and insufficient skills in cyber security and the importance of cyber hygiene from a young age hence the need to promote cyber security through education. France was time specific to include cyber security in the curriculum at the start of the 2016 academic year, stating education and training will be provided to different stakeholders such as employees, businesses, civil servants and teachers in higher education. Likewise, Estonia's lifelong learning program (Ministry of Education and Research, 2014) gives a breakdown of how cyber security will be integrated into the society through formal and informal education. National cyber security exercises in Lithuania run regularly to improve cyber security skills and promote cyber safety. Although four universities in Lithuania offer cyber security programmes, there is still a huge demand for cybersecurity specialists, therefore, necessitating the need for more study programs to be established. In Singapore, youths are being educated via cyber-wellness messages by the Inter-Ministry Cyber Wellness Steering Committee and the Cybercrime prevention awareness will be intensified in schools through Collaborative Social Programme (CoSP). Universities have integrated cybersecurity programmes for those interested in pursuing a cybersecurity education. For instance, Masters in Cybersecurity can be pursued at the Singapore University of Technology and Design (SUTD).

The UK strategy addresses skills shortage through training and education across various levels such as introducing cyber security skills into the education system, balancing the gender gap in cyber related professions, providing training and education programmes for 14–18-year-olds among other initiatives. The Department of Education in England invested £84m in promoting computing skills in schools (HM Government, 2018) which will promote the understanding of the subject area (Sentance, 2019). The National Audit Office (NAO) unfortunately could not measure the impact of this activity (National Audit Office, 2019). The Work-Integrated Learning (WIL) program launched in Canada as stated in its action plan includes 1,000 placement opportunities over three years in cyber security and the evaluation of the program will be conducted in 2021 (Public Safety Canada, 2019). Norway and Spain did not refer to actions in place and the actions in the US are generic compared to the other countries.

*Cybersecurity Research and Development (R&D) programmes:* Cybersecurity research and development programmes looks at software or hardware solutions (ITU, 2019). France takes a scientific and technical approach to cybersecurity and information protection by creating an expert panel for a digital trust whose objective is to discover key technologies, monitor and support research and provide support to PhD holders. Large equipment used for electronic communication networks is primarily developed outside the country and Europe creating a trust issue in the digital economy. It however aims to overcome this challenge by creating an environment that will support and promote its digital products nationally and internationally. Australia and Estonia also rely heavily on overseas IT products and services with Estonia's products mainly from Asia and the US which means that the risk associated with such IT solutions are also inherited thereby impacting their cyber security situation. Products and services developed in the country will be promoted at international level. Table 9 below shows some R&D programmes across the strategies.

Table 9: R&D Programmes

| Country | Actions |
|---------|---------|
| CAN | • Pan-Canadian Artificial Intelligence Strategy for research and talent.<br>• Quantum computing and blockchain technologies |
| SG | • SUTD Cyber Security Laboratory. |
| NOR | • No R&D programs outlined. |
| USA | • Promote 5G technology and lay a foundation to advance innovation. |
| ESP | • Improve research and development programmes in universities, businesses and research centres. |
| UK | • Establish two innovation centres to foster research and development and invest in defence and security. |
| LTH | • Invest in R&D through the EU research and innovation programme Horizon 2020.<br>• Motivate the private sector to develop tools and services in the cyber security area. |
| EST | • Research measure was set up in 2018 to foster research and to define R&D areas in cybersecurity.<br>• Cooperation between research, industry, Government, and academia will be promoted. |
| AUS | • Cyber Security Growth Centre to foster collaboration between the research community, start-ups, businesses, and the Government. |
| FRA | • The Government will create an environment that will foster research and development. At the time of documentation, requests were made for proposals to develop equipment that can detect cyber-attacks. |

*Incentive mechanisms:* this looks at incentives provided to support capacity building in cyber security. Incentives boosts defences against cyber threat through increased demand for cyber security products and services. Australia's incentive mechanism is robust and extensive compared to the other countries ranging from tax breaks to scholarships as shown in table 10.

Table 10: Incentive measures

| Country | Actions |
|---|---|
| CAN, NOR, LTH | • No incentive mechanism mentioned |
| FRA | • Products and services with cyber security risk analysis will receive a bonus factor when requesting for public funds, public projects or answering a call for bids. |
| SG | • Support home-made solutions and make it easier for cybersecurity companies to enter the market. |
| USA | • Provide incentives to cyber security investments, but no specific ventures mentioned. |
| ESP | • Provide incentives to support innovation particularly in small businesses. |
| UK | • Support start-ups, and provide testing facilities for companies |
| EST | • Set up mentoring programmes, seminars and events. |
| AUS | • Support cyber security start-ups<br>• Promote Data61 – the Commonwealth Scientific and Industrial Research Organisation's (CSIRO) digital powerhouse and providing PhD scholarship in cyber-related program.<br>• Concessional tax treatment to support start-ups; changes to tax treatment to attract investors; attract entrepreneurs by enhancing the visa system. |

**4.12 Cooperation measures**

The problems posed by cybercrime is a global one and necessitates stakeholders both at national and international levels to collaborate and this can be achieved in different ways such as international fora, bilateral and multilateral agreements and public-private partnership among others (ITU, 2019) as highlighted below:

*International cooperation:* the global internet is sustainable with the right balance between freedom, security, openness and robustness (Norwegian-Ministry-of-Foreign-Affairs, 2017). All the countries except Canada explicitly expressed existing or future action plans in place to promote global collaboration and made reference to international bodies such as INTERPOL (Singapore), NATO (Norway, Spain, UK, Lithuania, Estonia), UN (Norway, Spain, Lithuania, Estonia), EU (Norway, Spain, UK, Lithuania, Estonia, France), OECD (Norway), OSCE (Norway, UK, Lithuania, France), Commonwealth (UK) and ASEAN forum (Australia, Singapore). The USA strategy however lacks reference to specific activities with international partners or involvement with other nations to promote cyber security. Four guiding principles underpin Norway' international cooperation with an emphasis on cooperation and handling cyber incidents while Estonia refers to sustainability and security of digital society as the foundation on which international leadership and research and development thrives. All the countries except Singapore and Australia are part of the North Atlantic Treaty Organisation (NATO, 2019). Across the strategies, there is a consensus that international cooperation is key to a secure cyberspace shown in table 11.

Table 11: International Cooperation Measures

| Country | Bilateral and Multilateral Cooperation |
|---|---|
| CAN | • Bilateral and multilateral agreements with foreign governments not mentioned |
| SG | • Member of the Association of Southeast Asian Nations (ASEAN) which enables the members to share information and work in partnership<br>• Partners with INTERPOL and other member countries to address cybercrime. |
| NOR | • Bilateral collaborations with Nordic countries and multilateral agreements with organisations such as North Atlantic Treaty Organisation (NATO), United Nations (UN), the European Union (EU), the OECD and the OSCE. |
| US | • Strengthen cooperation with the United Nations and other partners. |
| ESP | • Multilateral and bilateral cooperation with organisations such as the United Nations, European Union, Atlantic Alliance and third-party states. |
| UK | • Bilateral and multilateral relationships with organisations such as NATO, UN, G20, OSCE, Council of Europe, the Commonwealth and the EU. |
| LTH | • Will pursue bilateral and multilateral cooperation with the EU, NATO, the United Nations, the Organisation for Security and Co-operation in Europe (OSCE), organisations of the Baltic region, USA and other countries. |
| EST | • Bilateral and multilateral cooperation with EU, NATO, OSCE and UN. |
| AUS | • Cooperation with New Zealand, East Asia Summit and Asia-Pacific countries. |
| FRA | • Cooperation with the European Union Agency for Network and Information Security (ENISA), NATO NCIRC (Computer Incidence Response Capability) and CERT-EU (Computer Emergency Response Team of the European Union (EU). |

*Public-Private Partnerships (PPP)*

All the countries express the significance of both public and private sectors working in partnership which involves sharing and exchange of information. Table 12 shows some key actions in the strategies.

Table 12: Public-Private Partnership

| Country | Key actions |
|---|---|
| CAN, NOR, EST and FRA | • The importance of PPP was noted but no reference to current or future actions. |
| SG | • The Partnership for the Advancement of the Cybersecurity Ecosystem (PACE) and ST Electronics-SUTD Cyber Security Laboratory is PPP initiatives. |
| US | • The Government will work in collaboration with the private sector to protect critical infrastructures. No specific actions mentioned. |
| ESP | • Set up of the National Cybersecurity Forum to strengthen collaboration between the public and private sector. |
| UK | • Support collaboration among the industry, academia and the Government by working with research councils and Innovate UK and industries such as CyberInvest will be established to aid such collaborations. |
| LTH | • The Government set up the Cyber Security Council to promote PPP using the Cyber Security Network to ensure the PPP. |
| AUS | • The establishment of joint cyber threat sharing centres will enable the Government and the private sector work in partnership to exchange information. The academic, business, government and research professionals work in partnership to run an annual cyber security competition for tertiary students. |

## 5    Discussion

One of the steps taken by countries to address the challenges of cyberthreat is developing a national cybersecurity strategy which describes the urgency of the cybersecurity task, initiatives to be employed, responsible agency for implementing strategy and the goals of the strategy (Dedeke & Masterson, 2019). NCSS provides a means through which a nation communicates its stance on cyber security reflecting the political, social, economic and financial climate of the nation. It also provides an opportunity for governments to communicate the economic benefits and associated risks of using the Internet while, simultaneously building the trust and confidence of users, protecting data and critical infrastructures (Kilmburg, 2012; Oxford Analytica, 2013). Although these nations have prepared their strategies on an individual basis and have aligned their socio-economic goals in various ways such as digital innovation (Canada, Norway, USA, Lithuania, UK, Australia), InfoComm technologies (Singapore), digital autonomy (Spain, France) and digital technology (Estonia), they, however, have a common goal to promote research and development by providing support to businesses, encouraging cybersecurity and protecting critical infrastructures. They also have similar strengths such as promoting international cooperation, public-private partnership, capacity building, research and development among others. A common strength is that eight out of the ten nations (AUS, CAN, NOR, ESP, USA, UK, FRA AND EST) have developed more than one strategy they can build upon, but, only UK, CAN and ESP referred to building on the accomplishments of previous strategy while EST referred to building on lessons learned. While some accomplishments were noted, none of the countries highlighted the failings of previous strategies as a steppingstone to developing an effective one. An essential element in improving an organisation's processes is looking at lessons learned, thereby necessitating storage and retrieval of useful lessons (Banerjee et al., 2019). Since governments play a major role in the cyber security of their nations, a strong inter-ministerial cooperation becomes vital as noted across the strategies. A good way of showing the connectivity of government ministries is by designing an organisational structure which according to Ahmady et al. (2016) provides the means through which the activities in an organisation are coordinated, organised and divided. Only ESP, EST and AUS showed an organisational structure for cyber security with EST showing a well laid structure which other strategies can emulate. Furthermore, five fundamental weaknesses common to the strategies were noted, which are discussed below.

## 5.1 Unreported methodology

A common weakness across the NCSSs is the unreported method and methodology in developing the strategy. The methodology process of any given field is an important aspect in the design of a strategy (Dragnea, 2016) as it transcends research and is applicable to government sectors providing a framework for objectives, methods and results thereby showing the diligence and vigour put into the work (Stuart, 2017). Though used interchangeably at times, there is however a difference between method and methodology. Methods are "ways of collecting data (such as observations, surveys and interviews), and methodologies, the principles and understandings that guide and influence our choice and use of methods (like experimentation and ethnography)" (Hyland, 2016).

Various methodologies have been developed for national strategies such as the national intellectual property strategy where a methodology, WIPO, was developed with the aim of providing a uniform and consistent approach to strategies and policies and also provide guidelines to those involved in preparing the strategies (WIPO, 2016). Another example is the methodology designed for radioactive waste and spent fuel management strategies of Nuclear Energy Agency (NEA) member countries with the aim of having a uniform approach in presenting data on spent fuel and radioactive waste (OECD, 2017). A further example is the Project IARM (Identifying and Assessing the Risk of Money laundering), a seven-step methodological approach designed to evaluate the risk of money laundering which was tested in three countries – UK, Italy and the Netherlands (Savona & Riccardi, 2017). While UK, SG, US, LTH, FRA and AUS failed to state their method and methodology, CAN, NOR, ESP and EST vaguely stated workshops, conferences, consultations, evaluations and stakeholder inclusiveness as a form of method and methodology.

## 5.2 Absence of / inadequate implementation and evaluation plan

Another weakness is the absence of implementation and evaluation plan. "Strategy implementation refers to the mechanisms, resources, and relationships that help to translate the strategy into action" (Platt et al., 2019, p.) however, there are challenges that can hinder successful implementation such as limited or inadequate allocation of resources, lack of cooperation among personnel and poor monitoring of progress (Platt et al., 2019). In a guide developed to evaluate Marine Spatial plans, it was noted that an evaluation plan should also be developed along with the strategy (Ehler, 2014). The guide identified eight steps of monitoring and evaluating the performance of marine spatial plans with each step having tasks and outcome. Four benefits of performance monitoring and evaluation noted were accountability and transparency, knowledge and information, learning and political and public support (Ehler, 2014).

The World Health Organisation (WHO) guide for developing a national Health Financing Strategy (HFS) note that an evaluation plan is fundamental to the HFS and should not be an afterthought Kutzin et al. (2017). It further noted that while monitoring progress is important, it is however not sufficient, hence the need for evaluation. In the review of the HFS of eight developing countries, Cali et al. (2018) noted that none of the countries had a well-developed evaluation strategy but rather designed their strategies as an ad hoc or one-off task.

Four nations (USA, SG, ESP and FRA) lack both implementation and evaluation plan within their strategies; UK has a scanty implementation program within the strategy as it failed to set the strategic outcomes on a short–long term basis, failed to allocate outcomes to responsible unit and lacks an evaluation plan; Estonia's implementation plan was spread across the documentation with key performance indicators to measure its success. Australia's strategy has an action plan, but lacks an implementation and evaluation plan; Canada has a well-structured action plan in a separate action plan document (Public Safety Canada, 2019 ), it however lacks an evaluation plan; Norway's list of measures for its NCSS also lacks an evaluation plan; Lithuania on the other hand has an evaluation plan in accessing the success of its strategy.

## 5.3 Irregular assessment of strategy

A further weakness is the irregular assessment of the strategy. Only four nations (ESP, LTH, UK, AUS and EST) referred to an annual report or interim assessment of strategy, however, UK failed to produce annual reports due to sporadic governance (National Audit Office, 2019). Australia on the other hand produced only one annual report in 2017 (Australian Government, 2017) since the development of its strategy in 2016. The update indicates a gradual cultural change and increased partnership between Government, businesses and academia; however, the strategy has been criticised for spasmodic governance (Tonkin, 2019) and lack of support given to Australia' corporate sector (Phair, 2019). Although the update gives an overview of the achievements so far and future steps, it failed to identify the failures, challenges, and any recommendations. It is too early to ascertain whether Spain, Lithuania and Estonia will indeed produce an annual update.

## 5.4 Lack of transparency on allocated budgets

Organisations allocate budget and although it varies depending on the size and nature, it is however important for the smooth running of the organisation (Tang, 2009). Most of the strategies failed to state allocated funding for the implementation of the strategy which could mean that funding was done on an ad hoc basis. Only three nations (AUS, CAN and UK) stated the allocated budget for the strategy, but the Australian strategy has been criticised for inadequate funds (Phair, 2019; Dorian, 2019). The UK strategy lacked a business case; hence it is uncertain whether the strategy was adequately funded (National Audit Office, 2019). It is not clear if Canada's strategy was adequately funded. Moreover, five nations (NOR, SG, USA, FRA and ESP) did not give a timeline for the implementation of their strategies which means that it may be difficult to determine if the goals of the strategy will be achieved.

## 5.5 Absence of framework

Furthermore, a common weakness is the absence of a framework for developing the strategies. Although there is no universal framework for developing a national cybersecurity strategy, there are frameworks that have been developed for national strategies and organisations to manage cybersecurity. One of such is the NIST (National Institute of Standards Technology) cybersecurity framework which consists of three parts: the framework core, framework implementation tiers and framework profile (Sedgewick, 2014). Each part has its embedded components, for example, the framework core is made up of four components: functions, categories, subcategories, and Informative references, each with a set of activities. Another framework is the PCI DSS (Payment Card Industry Data Security Standard) which regulates how debit and credit card information are handled and consists of twelve requirements, categorised into six control objectives that organisations must comply with (Ukidve et al., 2017; Calder & Williams, 2019).

Different organisations have developed framework for national cybersecurity such as the Commonwealth Approach for Developing National Cybersecurity Strategies (CTO, 2015), Microsoft's Developing a National Strategy for Cybersecurity (Goodwin & Nicholas, 2013), ITU National Cybersecurity Strategy Guide (Wamala, 2011), NCSS Good Practice Guide (ENISA, 2016), National Cybersecurity guidelines (NATO, 2013), National Cyber Security Strategies Practical Guide on Development and Execution (European Network and Information Agency, 2012) and Guide to Developing a National Cybersecurity Strategy (ITU, 2018). Framework for national strategies have also been developed in other fields such as the WHO practical approach for developing policy and strategy to improve quality care (WHO, 2018) recognising that countries differ in various ways, however, there are similarities in the effort to improve quality care. The framework consists of eight elements and an operational plan for implementation.

The framework used to in mapping the ten NCSS as shown in figure 2 is the "A Guide to developing a national cybersecurity strategy: strategic engagement in cybersecurity developed by twelve partners from intergovernmental and international organisations, private sector, academia and civil society" (ITU, 2018). Although some of the NCSSs were developed before the guide, it however provides a platform that nations can use to improve on future strategies. The mapping is centred around two sections of the guide: Overarching principles for a strategy (nine principles to be considered when developing NCSS) and National Cybersecurity good practices comprising of seven focus area. The overarching principles if integrated can help produce a progressive and comprehensive national cybersecurity strategy while the good practice can produce an effective strategy.

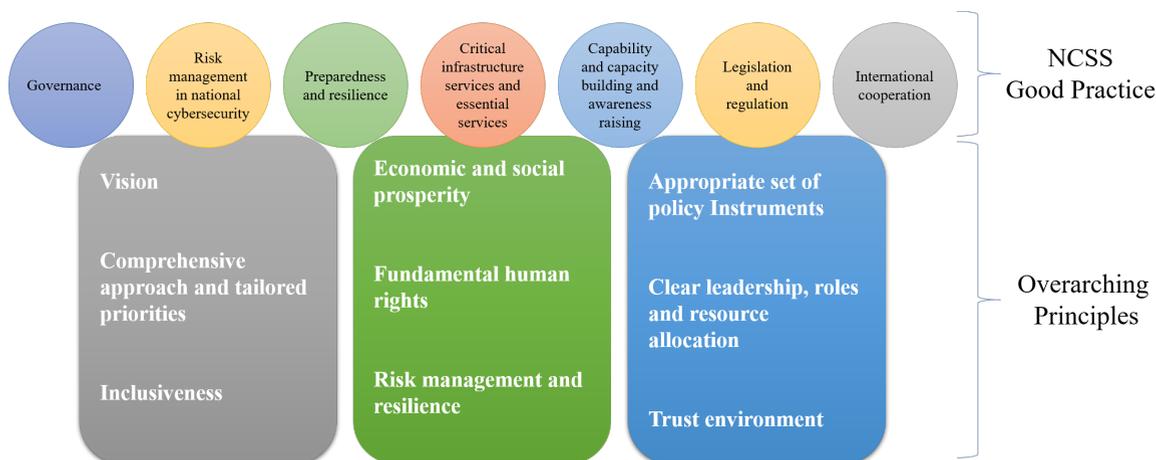

Figure 2: Overarching principles and National Cybersecurity Strategy Good Practice to developing a National Cybersecurity Strategy. Adapted from: Guide to Developing a National Cybersecurity Strategy – Strategic engagement in cybersecurity, (ITU, 2018)

## 6 Framework mapping of NCSS

The vision section of the guide above has been adapted for the purpose of this research and split into vision, objectives, scope and timeline for a comprehensive outcome as shown in table 13. The first section, overarching principles, show common strength in stating the vision, objectives, priorities, human rights, policy instruments and trust. All the countries except FRA stated their vision. The likelihood of a strategy being successful depends on a clear vision, the goals, the rationale and scope (ITU, 2018). In ensuring the achievement of the goals and objectives, a strategy should have a set timeline and reflect the priorities, beliefs and cultural values of the nation (Goodwin & Nicholas, 2013; ITU, 2018). Common weak areas in the overarching principles are scope, timeline, inclusiveness, clear leadership, roles and resource allocation. Seven countries (SG, NOR, US, ESP, LTH, AUS and FRA) failed to state the scope of their strategies while five countries (SG, NOR, US, ESP and FRA) failed to give a timeline for their strategies. Also, the involvement of stakeholders in developing and implementing the strategy is imperative to the success of the strategy, as Government's failure to identify and foster cooperation with other stakeholders limits the influence of the strategy (Kilmburg, 2012). Six countries (SG, US, UK, LTH, AUS and FRA) did not state stakeholder participation in the development of the strategy. Furthermore, the roles and responsibilities of stakeholders should be clearly defined to ensure active participation and implementation of the strategy. Across the strategies, governments play a major role in ensuring cybersecurity, however, allocation of financial, human and material resources is vital for the implementation of the strategy (ITU, 2018). Four countries SG, LTH, FRA and UK have a single agency for national cybersecurity, but none have statutory powers. Three countries (UK, CAN and AUS) have a dedicated budget for strategy and just three countries (EST, NOR, ESP) summarized the roles of government ministries involved in cybersecurity.

The second section, national cybersecurity good practice, is set across seven focus areas: governance, risk management in national cybersecurity, preparedness and resilience, critical infrastructure services and essential services, capacity and capacity building and awareness raising, legislation and regulation, international cooperation. In structuring an effective national cybersecurity, a Microsoft publication gave five recommendations: appoint a single national cybersecurity agency; provide the national cybersecurity agency with a clear mandate; ensure the national cybersecurity agency has appropriate statutory powers; implement a five-part organisational structure and expect to evolve and adapt (Nicholas et al., 2017). In the governance focus area, all the NCSSs show support from the highest authority, the Government, which indicates that cybersecurity is of high importance to the economic and social wellbeing of the countries but only six countries (SG, US, UK, LTH, EST, AUS) explicitly stated a competent cybersecurity authority.

Common weak areas in the section are poor intra-government cooperation, lack of allocation of dedicated budget and lack of implementation plan. Three core functions: commitment, coordination and cooperation are central to inter-government cooperation, however, the NCSSs have not particularly displayed strong emphasis across the three functions. Communication and coordination ensure governmental institutions understand their roles and interconnectivity while commitment is supporting policies to ensure delivery of the strategy. As earlier noted, only three countries, UK, CAN and AUS stated dedicated budget for the implementation strategy but the National Audit Office report (2019) reveals that a delay in the recruitment of three personnel to promote British cyber security companies abroad was due to a lack of agreement on the terms of the role (National Audit Office, 2019). In Australia, new roles were created: Minister assisting the Prime Minister, special adviser on cyber security, and Cyber Ambassador. A report by Tonkin (2019) reveals that there was an eight-month delay in the appointment of the Cyber Ambassador inevitably hampering the progress of the strategy at the time. Unfortunately, a change in Government in 2018 led to the removal of the post of Minster assisting the PM on cyber security and was reassigned to the Department of Home Affairs. One of the recommendations of the previous Australian strategy (Public Safety Canada, 2017) was to ensure governance committees have clearly defined roles and responsibilities as there were overlap duties in the roles of some of the committees. Although the subsequent action plan allocated the objectives to responsible departments but, there were no performance measurements with the plan.

While risks cannot be eliminated completely, they can however be managed. In the focus areas of risk management and critical infrastructure services, the NCSSs failed to coherently define a risk management approach that identifies vital services, threats, and associated risks stating the responsibilities of the main entities in each sector. Also, the risk management approach was not based on international standards neither were certification programs identified. Furthermore, NCSS should recommend sectoral cybersecurity risk profiles for vital sectors of the economy as it not only provides support for individual risk assessment in organisations, but also reduces the resources organisations need for risk assessments (ITU, 2018); these risk profiles should however be updated regularly.

Table 13: Mapping of NCSSs      º = present     ×= absent     ◊ = not clearly expressed

| | CAN | SG | NOR | US | ESP | UK | LTH | EST | AUS | FRA |
|---|---|---|---|---|---|---|---|---|---|---|
| **Overarching principles** | | | | | | | | | | |
| Vision | º | º | º | º | º | º | º | º | º | × |
| Objectives | º | º | º | º | º | º | º | º | º | º |
| Scope | º | × | × | × | × | º | × | º | × | × |
| Timeline | º | × | × | × | × | º | º | º | º | × |
| Comprehensive approach and tailored priorities | ◊ | º | º | º | º | º | º | º | º | º |
| Inclusiveness | º | × | º | × | º | × | × | º | × | × |
| Economic and social prosperity | º | º | º | º | º | º | º | º | º | º |
| Fundamental human rights | º | ◊ | º | º | º | º | º | º | º | º |
| Risk management and resilience | ◊ | º | ◊ | ◊ | º | ◊ | ◊ | º | º | ◊ |
| Appropriate set of policy Instruments | º | º | º | º | º | º | º | º | º | º |
| Clear leadership, roles and resource allocation | × | × | × | × | × | ◊ | × | × | × | × |
| Trust environment | º | º | º | º | º | º | º | º | º | º |
| **National Cybersecurity Strategy Good Practice** | | | | | | | | | | |
| **Focus area 1 – Governance** | | | | | | | | | | |
| Ensure the highest level of support | º | º | º | º | º | º | º | º | º | º |
| Establish a competent cybersecurity authority | × | º | ◊ | º | ◊ | º | º | º | º | × |
| Ensure intra-government cooperation | × | ◊ | ◊ | ◊ | ◊ | ◊ | ◊ | ◊ | ◊ | ◊ |
| Ensure inter-sectoral cooperation | × | º | º | ◊ | ◊ | º | ◊ | º | º | º |
| Allocate dedicated budget and resources | × | × | × | × | × | ◊ | × | × | × | × |
| Develop an implementation plan | × | × | × | × | × | ◊ | º | º | × | × |
| **Focus area 2 – Risk management in national cybersecurity** | | | | | | | | | | |

| | CAN | SG | NOR | US | ESP | UK | LTH | EST | AUS | FRA |
|---|---|---|---|---|---|---|---|---|---|---|
| Define a risk management approach | × | × | × | × | × | ◊ | × | × | × | × |
| Identify a common methodology for managing cybersecurity risk | × | × | × | × | × | ◊ | × | × | × | × |
| Develop sectoral cybersecurity risk profiles | × | o | × | × | × | × | × | × | × | × |
| Establish cybersecurity policies | × | o | × | o | o | o | o | o | o | o |
| **Focus area 3 – Preparedness and resilience** | | | | | | | | | | |
| Establish cyber-incident response capabilities | o | o | o | o | o | o | o | o | o | o |
| Establish contingency plans for cybersecurity crisis management | × | o | ◊ | ◊ | ◊ | o | ◊ | ◊ | × | × |
| | | | | | | | | | | |
| Promote information-sharing | o | o | o | o | o | o | o | o | o | × |
| Conduct cybersecurity exercises | × | o | o | o | o | o | o | o | o | o |
| **Focus area 4 – Critical infrastructure services and essential services** | | | | | | | | | | |
| Establish a risk-management approach to protecting critical infrastructures and services | × | o | × | × | × | × | × | × | × | × |
| Adopt a governance model with clear responsibilities | × | o | × | × | × | o | × | × | × | × |
| Define minimum cybersecurity baselines | × | o | × | × | × | o | × | o | o | o |
| Utilise a wide range of market levers | × | o | × | o | × | o | × | × | × | × |
| Establish public private partnerships | o | o | o | o | × | o | o | × | o | o |
| **Focus area 5 – Capability and capacity building and awareness raising** | | | | | | | | | | |
| Develop cybersecurity curricula | o | o | o | o | o | o | o | o | o | o |
| Stimulate skills development and workforce training | o | o | o | o | o | o | o | o | o | o |
| Implement a coordinated cybersecurity awareness-raising programme | ◊ | o | ◊ | ◊ | ◊ | o | × | o | o | o |
| Foster cybersecurity innovation and R&D | o | o | o | o | o | o | o | o | o | o |
| **Focus area 6 – Legislation and regulation** | | | | | | | | | | |
| Establish cybercrime legislation | ◊ | o | × | o | o | o | o | o | o | o |
| Recognise and safeguard individual rights and liberties | o | o | o | o | o | o | o | o | o | o |
| Create compliance mechanisms | × | o | × | o | o | o | × | o | o | × |
| Promote capacity-building for law enforcement | o | o | o | o | o | o | o | o | o | o |
| Establish inter-organisational processes | o | o | o | o | o | o | o | o | o | o |
| Support international cooperation to combat cybercrime | × | o | o | o | o | o | o | o | o | o |
| | CAN | SG | NOR | US | ESP | UK | LTH | EST | AUS | FRA |
| **Focus area 7 – International cooperation** | | | | | | | | | | |
| Recognise the importance of cybersecurity as a priority of foreign policy | o | o | o | o | o | o | o | o | o | o |
| Engage in international discussions | × | o | o | × | o | o | o | o | o | o |
| Promote formal and informal cooperation in cyberspace | × | o | o | o | o | o | o | o | o | o |
| Align domestic and international cybersecurity efforts | × | o | o | o | o | o | o | o | o | o |

Across the NCSSs, not much reference was made to sectoral risk profile except SG which categorically stated the design of cybersecurity into the financial sector through the establishment of a Financial Technology and Innovation Group with the responsibilities of developing regulatory policies and strategies for risk management and use of technology to increase effectiveness in the sector. Additionally, in the risk management approach to protecting critical infrastructures, only SG discussed extensively on their systematic cyber risk management approach, providing a framework which includes identifying Critical Information Infrastructures (CIIs), the use of a maturity index to access the readiness of critical infrastructure owners, audit performance and awareness building, raising professional standards through certifications and accreditations among others. While most of the NCSSs have an incident response plan and aim to improve on existing ones, SG however discussed its contingency plans for crisis management which consists of three levels of response to threats at national, sectoral and specific operator levels.

In improving cyber hygiene, all the NCSSs acknowledged the need to invest in capacity building by encouraging re-skilling, introducing cybersecurity into education, promoting professional qualification and working with different stakeholders to improve awareness but only five countries (SG, UK, EST, AUS and FRA) explicitly stated responsible authority for coordinating awareness programs. Having a competent authority that coordinates awareness programmes ensures efficient use of resources and accountability. Although the guide is comprehensive, it however failed to state the methodology and model used in developing the guide. While there were references to each recommendation, it is however beneficial to include practical steps or tasks strategy developers can take to accomplish the recommendations.

Although there is no unified framework for developing a national cyber security strategy, developers of cybersecurity strategy should consider three factors of utmost importance to cybersecurity – people, processes and technology. The gap between technology and people is addressed by processes which includes policy and procedures that should be included in the training and awareness program to provide instructions that should be followed in certain circumstances (LeClair et al., 2013; Janes, 2019). Processes should be properly documented, reviewed regularly and be current to ensure its effectiveness while technology on the other hand can be used in enforcing polices, providing data protection and monitoring and alerting violations (Janes, 2019). The risk of data loss either by human error or deliberate action can also be addressed through technological solutions (Janes, 2019).

No single factor can mitigate the challenges of cyber threats (Cook et al., 2011), therefore, it is imperative to understand the relationship between these factors and their roles in cyber security as prioritising one or two of these factors and neglecting the other increases vulnerability. Technology, interaction of processes and users' responses to major security events are all important considerations of cyber security (Bowen et al., 2011). For instance, in determining how secure an organisation is, it is imperative to know if employees are well educated and trained, if organisations comply with regulations on protecting and managing data and how to measure the security risk on new technologies or services provided (Stolfo et al., 2011) A system can become vulnerable due to a mistake in the design either deliberately or an accidental omission which questions the systems integrity, confidentiality and availability (Kostopoulos, 2017) and these vulnerabilities can be minimised by paying attention to people, processes and technology (Janes, 2019). As people, processes and technology are the backbone of cyber security, a robust national cyber security strategy (NCSS) should therefore be linked to these factors.

# 7 Conclusion

A strategy communicates the direction of an organisation on a long-term basis (Johnson et al., 2014) and the comparative analysis of the national cybersecurity strategies of the selected countries showed that considerations were given to the contents of the documents and the mapping based on the framework showed that the contents expected in NCSS were mostly present. In accomplishing the desired outcomes of the strategy, it is vital for government ministries to work in partnership, promote

Public Private Partnership (PPP), allocate appropriate funding and have an effective implementation plan. Leadership roles and responsibilities should also be clearly defined, whilst considering the material, human and financial resources needed (ITU, 2018). Common overarching principles across the strategies are objectives, comprehensive approach and tailored priorities, inclusiveness, economic and social prosperity, appropriate set of policy instruments and trusted environment. The strategies have been hugely supported at the highest level and there is an understanding that cyber-attacks are inevitable, and Government alone cannot drive cyber security, hence the need for collaboration with the private sector especially the owners of critical infrastructures. There is also the need to have adequate plans in place to prevent attacks and recover in the event of any attack.

Furthermore, capacity building and raising of awareness of cyber security were well noted as important factors to the success of the strategy and the NCSSs expressed the need to invest in awareness, training, and education to promote cyber hygiene, address the shortage of cyber security skills and foster research and development. Moreover, in designing and developing a NCSS, nations should identify gaps in the national framework and develop lines of action to overcome the gaps in the area of policy, regulation, legislation and the roles and responsibilities of stakeholders (NATO, 2013) and these will vary across countries. It is observed that in legislation and regulation, cybersecurity is taken seriously as laws and processes are being reviewed to ensure they are fit for purpose. Law enforcement personnel are also being supported to ensure they can deal with cyber security issues, but more should be done in promoting compliance.

However, key factors which are critical to the success of a strategy such as implementation plan, evaluation plan, allocation of resources, risk management and annual assessment of strategy were seen to be either absent or inadequately expressed. Despite having similar viewpoints such as cybersecurity being a shared responsibility and the importance of working together nationally and internationally, there are differences in the understanding of the term cybersecurity. The term was used in all the strategies but only three nations defined the term, however, there was no unified definition. Some challenges such as overlap/lack of clarity in roles and responsibilities, shortage of cyber security skills, dependence on digitalisation and lack of allocation of resources were noted. While all the strategies have measures to protect the country, taking both offensive and defensive measures, UK however stated specific technical measures to improve cybersecurity across the country's network and the expected level of engagement with Communications Service Providers.

Due to the peculiarity of each nation, the level and modality of response to incidents will vary and this is seen in the different roles of the response teams. While some nations have different agencies that oversee cybersecurity in different sectors, some operate a one-stop shop principle for cybersecurity issues with only SG stating the level of authority their central agency has such as developing and enforcing policies and regulations. Of the five countries (ESP, EST, UK, LHU and FRA) that were part of the European Union at the time of developing the strategies, three countries (LTH, EST, ESP) developed their strategies after GDPR (General Data Protection Regulation) came into force but did not refer to it. UK however referred to using GDPR to drive cybersecurity standards in its strategy prior to it coming into force. Since the methodology of the NCSSs was either lacking or inadequate and the framework for developing the strategies was not stated, it is not clear if the NCSSs were meticulously prepared, or it was done on an ad hoc basis.

## 8 Recommendations

While nations will prioritise and develop strategies based on their needs, there are key areas that should be considered based on the findings of the comparative analysis for which recommendations are put forward below:

1. A NCSS should include an explicit definition of the term "cybersecurity" to convey its understanding of the term. A lack of or vague definition suggests little understanding and may lead to an omission of important aspects of cybersecurity. While there may not yet be a universal definition of the term, the developers of a NCSS should look beyond the current position of the nation and put into consideration emergent threats due to the dynamic nature of cybersecurity.
2. The goals of the strategy should be split into short, medium and long-term goals with responsible actors and realistic timelines in achieving them. Open-ended goals may lead to unfulfillment of the vision and objectives of the strategy.
3. The methodology for developing the strategy should be defined. This helps to ensure that possible approaches and framework to developing the strategy have been looked at and due consideration is given to the process.
4. Provision for annual progress report of strategy stating the achievements, challenges, and recommendations.
5. Implementation and evaluation plans should be developed along with the strategy.
6. The strategy should be based on an underlying framework to ensure it addresses every aspect of cybersecurity.
7. The risk management approach of strategy should be well defined.